\documentclass[reprint,
]{revtex4-1}

\usepackage{graphicx}
\usepackage{dcolumn}
\usepackage{bm}
\usepackage{amsmath}  
\usepackage{amsfonts} 

\begin{document}

\preprint{APS/123-QED}

\title{Probing higher OAM of LG beams via diffraction through translated single slit}

\author{Jadze Princeton C. Narag}
\email{jnarag@nip.upd.edu.ph}
\author{Nathaniel P. Hermosa}%
 \email{nhermosa@upd.edu.ph}
\affiliation{%
 National Institute of Physics, University of the Philippines, \\ Diliman, Quezon City Philippines 1101
}%




\date{\today}

\begin{abstract}
Laguerre-Gaussian (LG) beams carry orbital angular momentum (OAM) as a consequence of a singularity in their phase profile. In this work, we probe the OAM of LG beams using a movable single slit. We invoke the method of stationary phase to determine the resulting diffraction pattern for different topological charges, observe how these patterns changes when the slit is moved transversely and show that by looking at the resulting diffraction pattern for different slit positions, we can determine the orbital angular momentum of the incoming beam. Using a digital micromirror device (DMD) as a dynamic slit, we were able to access higher $\ell$ that were not possible in previous works involving slit diffraction. Our results are valuable in the study of OAM specially for beams where the current technology for detection is not yet mature, for example, beams whose wavelengths are in the terahertz regime or beams of higher energy such as, electron vortex beams.

\end{abstract}

\pacs{Valid PACS appear here}
\maketitle


\section{LG beams and the orbital angular momentum of light}
Light has been shown by Woerdman et al. \cite{allen1992allen} to carry orbital angular momentum (OAM) as a consequence of its spatial structure. In particular, light beams possessing an azimuthal phase dependence of $\exp(i\ell\phi)$ where $\ell$ is an integer, carry OAM of $\ell\hbar$ per photon. The canonical example of an OAM carrying beam is the Laguerre-Gauss (LG) beam with an electric field given by \cite{saleh2007fundamentals},
\begin{equation}
u_{\ell}^{p}(\rho,\phi,z) \approx \left(\frac{\rho}{\omega}\right)^{|\ell|} L^{|\ell|}_p \left(\frac{2\rho^2}{\omega^2}\right) \exp\left(\frac{-\rho^2}{\omega^2}\right)\exp\left(i\ell\phi\right)
\label{eq:LG}
\end{equation}
where $\omega$ is the beam waist, $L^\ell_p \left(\frac{2\rho^2}{\omega^2}\right)$ is the associated Laguerre polynomial, $\ell$ is the azimuthal mode, and $p$ is the radial mode. The discontinuity of the azimuthal phase leads to a singularity or vortex at the center of the beam. The azimuthal mode $\ell$, also called the topological charge, defines size of the vortex while $p$ defines the number of radial nodes. 

The OAM of the LG beam has led to its applications in quantum information \cite{molina2001management,torres2003preparation,straupe2010quantum}, micromachining \cite{ladavac2004microoptomechanical}, optical communication \cite{gibson2004free}, optical tweezing \cite{grier2003revolution}, and in astronomy \cite{foo2005optical}  and gravitational wave detection \cite{granata2010higher}. A common method to determine the OAM of an LG beam is through interferometry \cite{soskin1997topological,basistiy1993optics, basistiy1995optical, harris1994laser, padgett1996experiment}. Moreover, the diffraction of the beam through different apertures has been shown to also be practicable method of probing the OAM. For example, the diffraction of the LG beam through polygonal apertures \cite{silva2014unveiling, ambuj2014diffraction, hickmann2010unveiling, mourka2011visualization, de2011measuring} and multi-pinhole set-ups \cite{shi2012characterizing, berkhout2008method} are shown to be effective in probing their OAM . The classical single-slit \cite{ferreira2011fraunhofer, ghai2009single} and double-slit \cite{emile2014young, sztul2006double} diffraction also have been reexamined using LG beams instead of the ordinary Gaussian beam. However, in these works they used stationary slits and they were only able to probe the low ordered modes. In this paper, we study the diffraction of the LG beam through a dynamic single-slit using a digital micromirror device. We look at the fringe patterns formed and show that these are related to the topological charge of the LG beam. We explain the resulting structure of the diffraction using the method of stationary phase applied to the Huygens-Fresnel diffraction integral. Our results are valuable in the study of OAM specially for beams where the current technology for detection is not yet mature, for example, beams whose wavelengths are in the terahertz regime or beams of higher energy such as, electron vortex beams \cite{uchida2010generation, verbeeck2010production}.

\section{Slit diffraction of LG using stationary phase approximation}\label{sec:intro}
	The main assumption of the classical single slit experiment is that a plane wave impinges on the slit. That is, at the plane of the slit the waves are in phase. The resulting diffraction pattern can be derived by imagining secondary sources along the slit. The phase difference between these wavelets are caused by their path difference to the observation point. Mathematically, the field is given by the Huygens diffraction integral,
\begin{equation}
u = \int\int_S \frac{ik}{r} e^{ikr} dS
\end{equation}	
	where $r$ is the distance from the source point to the observation point, $k$ is the wavevector, and the integral is over the entire surface $S$ containing the source points. Applying the conditions for the Fraunhofer approximations and taking the intensity gives \cite{born2013principles},
	\begin{equation}
I \propto sinc^2(\beta)
\label{eq:sinc}
	\end{equation}
    where $\beta = kbsin\theta /2$, $b$ being the thickness of the slit and $\theta$ the angle from the center of the slit to the observation point. The sinc function means that there are alternating bright and dark fringes perpendicular to the slit and that the maximum intensity is found along the axis which contains about 85\% of the power transmitted. 

\begin{figure}[!h]
\centering
\includegraphics[width=\linewidth]{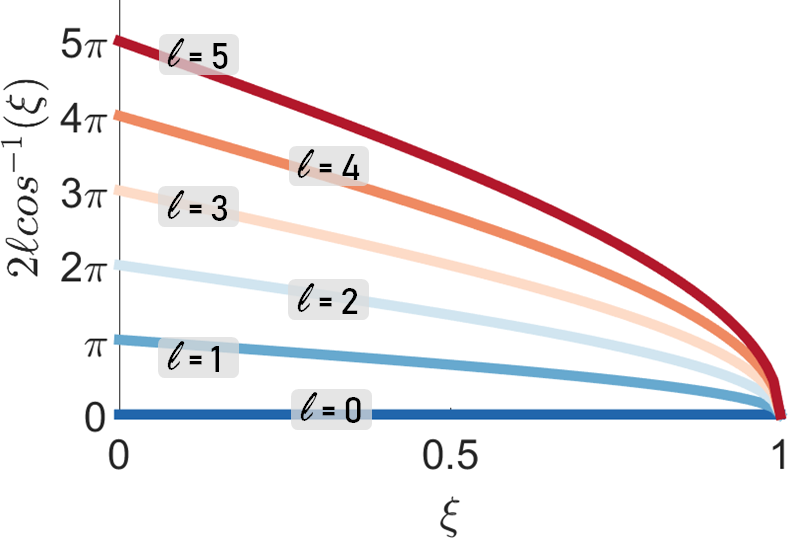}
\caption{$2 \ell cos^{-1}$ vs $\xi$. Here $\xi$ is a normalized distance, $\xi = \frac{x}{\sqrt{x^2+L^2}}$. The phase difference transitions from an odd to even (or even to odd) multiples of $\pi$ as the slit position $x$ becomes very large or as $\xi$ approaches 1. The number of transitions, from odd to even or even to odd multiples of $\pi$ depends on the OAM of the beam.}
\label{fig:cosineinverse}
\end{figure}

	If instead we use a higher order LG beam rather than a Gaussian beam, then the waves are no longer in phase in the plane of the slit. There is an additional phase difference due to the azimuthal phase dependence of the LG beam. This additional phase difference can be incorporated inside the exponential in equation 2. However, evaluating the integral can be a bit involved so instead we invoke the method of stationary phase. The method of stationary phase states that for an integral of the form,
    \begin{equation}
    \int g(z)e^{ikf(z)}dz,
    \end{equation}
    where function $f(z)$ changes rapidly compared to the wavelength in the domain of integration, the real and imaginary parts of the exponential oscillate rapidly and the contributions from each point in the domain of integration will cancel each other out except at the edge of the integration region where the cancellation is incomplete \cite{born2013principles,kathavate1945geometric}. For the case of a vertical single slit, this means that most of the contribution to the diffraction integral comes from the top and bottom endpoints. For the case of the plane wave, the sources from the top and bottom endpoints of a vertical single slit are in phase, and therefore will interfere constructively along the horizontal line along the center of the slit \cite{narag2018diffraction}. This is the ordinary pattern we see in the classical single slit diffraction. There is a bright band perpendicular to the slit. The horizontal fringe is not uniform but is a sinc function as in equation \ref{eq:sinc}. However for a very narrow slit we can approximate it to be a uniform band, because the width of the sinc function becomes wide.
    In the case of LG beams, the phase on the top and bottom edge of the slit will differ depending on the OAM and on the position $x$ of the slit with respect to the center of the beam. A simple calculation shows that the phase difference $\Delta \phi$ between the top and bottom edge of the slit is given by,
    \begin{equation}
\Delta \phi = 2 \ell \cos ^{-1} \left(\frac{x}{\sqrt{x^2+L^2}}\right)
\label{eq:delphi}
    \end{equation}
    where $\ell$ is the topological charge of the beam, $x$ is the displacement of the slit, and $2L$ is the length of the slit. We assume that the slit is very narrow. When the slit is centered, i.e. $x = 0$, equation \ref{eq:delphi} gives the phase difference to be $\Delta \phi = \ell \pi$. Thus, when $\ell$ is even the result is constructive interference and the central horizontal fringe is a bright band and when $\ell$ is odd, the interference is destructive and the central horizontal fringe is a dark band.  Moreover, $\Delta \phi$ also depends on the slit position $x$. Figure \ref{fig:cosineinverse} shows the trend of the $2\ell cos^{-1}$ as the slit is moved away from the center. When the slit is at the center of the beam, $\Delta \phi$ is an odd (even) multiple of $\pi$ for odd (even) $\ell$. As the slit moves away from the center, or equivalently as $cos^{-1}$ approaches 1, the phase difference decreases and goes to zero. The phase difference evolves from an odd to an even multiple of $\pi$ and this evolution depends on the topological charge $\ell$. For example, for $\ell = 1$ the phase difference will change from an odd to an even multiple of $\pi$ only once. For $\ell = 2$, $\Delta \phi$ starts as an even multiple of $\pi$ ($\Delta \phi = 2 \pi$) then changes to an odd multiple of $\pi$ ($\Delta \phi = \pi$) and finally back to an even multiple ($\Delta \phi = 0$) as the slit is moved. Thus, when the slit is moved the central horizontal intensity will transition from a dark band to a bright band only once for $\ell = 1$ while for $\ell = 2$ the central intensity will transition from a bright to a dark then again to a bright band. The number of transitions and how fast the transitions of the fringes from dark to bright (or bright to dark) depends on $\ell$. In general, the central fringe starts as a dark band for odd $\ell$ and a bright band for even $\ell$ and the evolutions of the fringes from dark to bright or vice versa is faster for larger $\ell$.

\section{Simulation}

\begin{figure}[h]  
\centering
\includegraphics[width=0.8\linewidth]{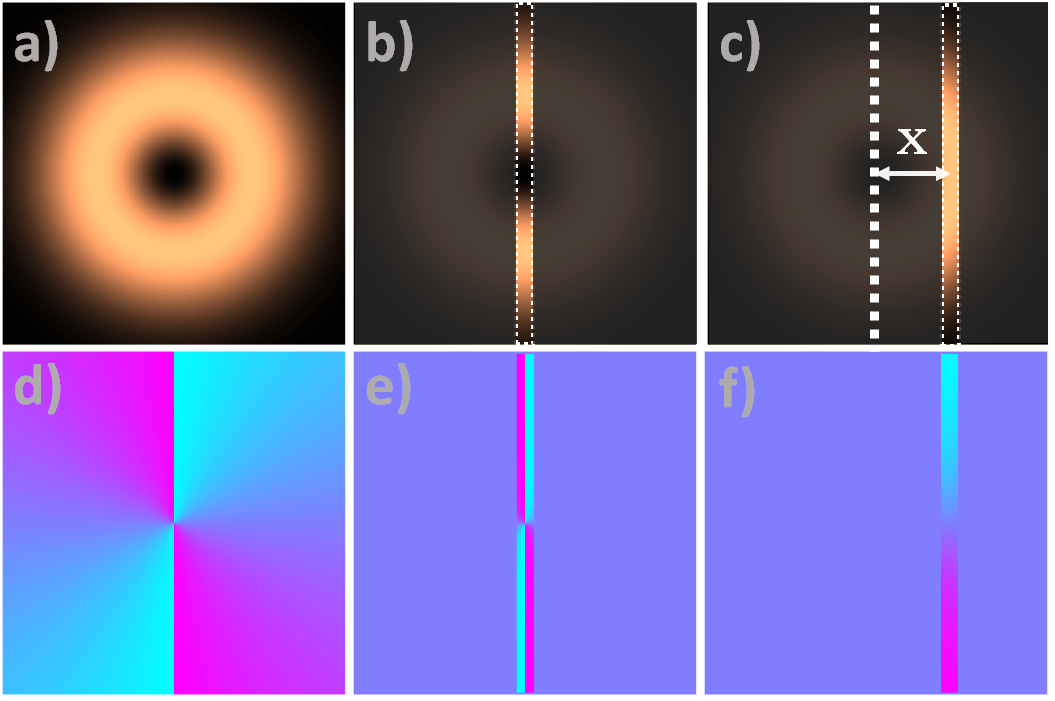}
\caption{Slit Modulation. Figure a and b show the intensity the $LG_2$ beam respectively before and immediately after the slit while shows intensity when the slit is moved at a distance x. d), e) and f) are the corresponding phase distribution.}
\label{fig:beforeandafterslit}
\end{figure}

We simulated the generation of $LG_\ell^{0}$ beams by approximating the beams using equation \ref{eq:LG}. The beams were then modulated in amplitude by a slit of thickness = 54 $\mu$m. This is done by multiplying the beam matrix element wise with the amplitude mask of the slit. The intensity of $|LG|^{2}$ beam (a) before the slit, (b) after the slit and (c) after the slit has been moved laterally are shown in Figure \ref{fig:beforeandafterslit}a to and \ref{fig:beforeandafterslit}c, respectively; while the corresponding phases are shown in  Figure \ref{fig:beforeandafterslit}d to Figure \ref{fig:beforeandafterslit}f. 

\begin{figure}[h]  
\centering
\includegraphics[width=\linewidth]{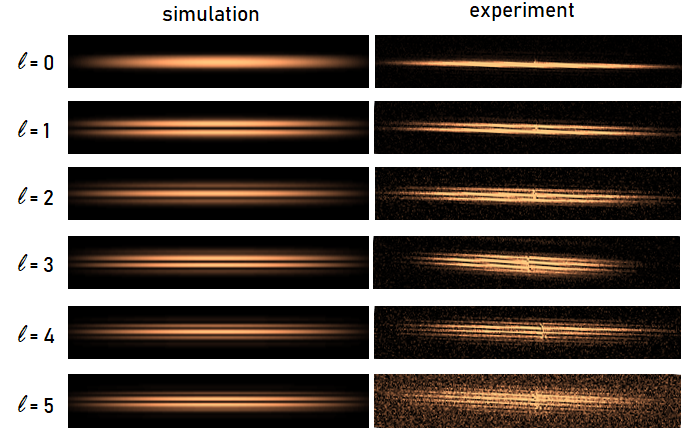}
\caption{Diffraction through a single slit of $LG_0^{\ell}$. For odd $\ell$ there is a central horizontal dark band while for even $\ell$ the central horizontal band is bright.}
\label{fig:DiffractionSimulation}
\end{figure}

The field right after the slit was then Fourier transformed to simulate the effect of the slit being placed in the back focal plane of a lens and being imaged in the front focal plane. The images for LG beams with different $\ell$ values and $x = 0$ are shown Figure \ref{fig:DiffractionSimulation}. The diffraction patterns show that there is an obvious difference between the diffraction pattern between LG beams with odd and even $\ell$. There is a horizontal dark line that cuts through the center of diffraction pattern for odd $\ell$ while for even $\ell$ the central horizontal band is bright. The number of fringes is also proportional to $\ell$. In particular the number of fringes is $\ell + 1$, same as the result in \cite{ambuj2014diffraction}. For higher $\ell$'s, however, it becomes more difficult to resolve the fringes. Also, the diffraction pattern becomes overall darker for larger $\ell$s since the vortex is also larger. That is why in the experimental result for $\ell = 5$ in fig. \ref{fig:DiffractionSimulation}, we had to increase the camera sensitivity to be able to capture the diffraction pattern and the images are noiser with more grains.

\begin{figure}[h]  
\centering
\includegraphics[width=0.9\linewidth]{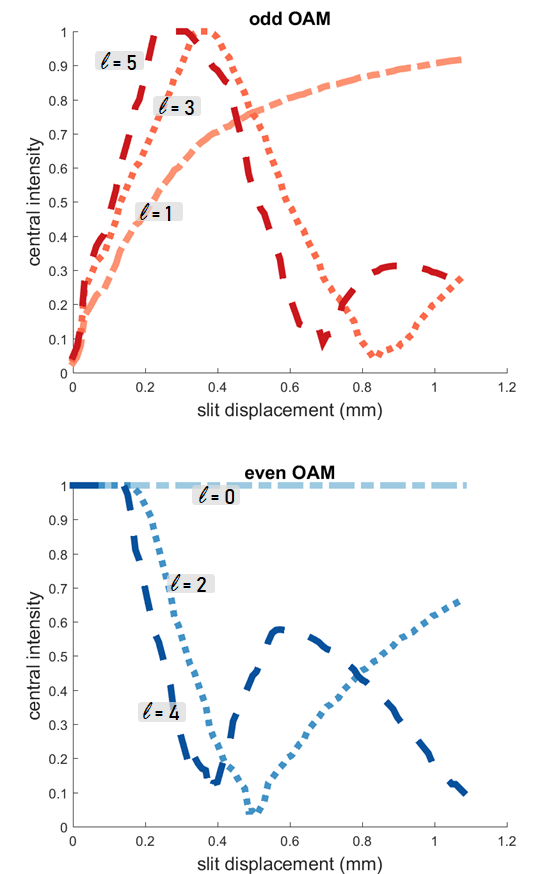}
\caption{Simulation of Central intensity vs slit position $x$. The central intensity for odd (even) $\ell$s start from a minima (maxima) and then changes to a maxima (minima) as the slit is moved from the center. Larger $\ell$s evolve faster than smaller $\ell$s.}
\label{fig:LineScanSimu}
\end{figure}

We also graphed in \ref{fig:LineScanSimu} the relative intensity of the central horizontal band as the slit is moved. Again, we see from the graphs that the central intensity when $x = 0$ for LG beams with odd (even) $\ell$s start at a minima (maxima). When the slit is moved the central horizontal band changes from an intensity maxima (minima) to a minima (maxima) for even (odd) $\ell$. This is explained by the variation of $\Delta \phi$ with respect to the slit position in equation \ref{eq:delphi}. The graph of equation \ref{eq:delphi} shows that for odd $\ell$, the phase difference starts as an odd multiple of $\pi$ and will alternately become an even then odd multiple of $\pi$ as the slit displacement becomes very large. For even $\ell$, the phase difference also alternates between odd and even multiples of $\pi$ but it starts, at $x = 0$, as an even multiple. This means that number of intensity change from maxima to minima (or minima to maxima) and how fast this change as a function of slit position are dependent of $\ell$. For example, for $\ell = 1$, the phase difference will change from an odd to an even multiple of $\pi$ only once. Thus the central intensity will only go once from a dark to a bright fringe. The rates are also faster for larger $\ell$. We see in figure \ref{fig:DiffractionSimulation} that for odd $\ell$ the central intensities are minima when $x=0$, but as we displace the slit the graph for $\ell = 5$ evolves faster than that of $l=3$ and that $\ell = 3$ is faster than that of $\ell =1$. The analogous behaviour is observed for even $\ell$'s: $\ell = 4$ evolves faster than $\ell = 2$ which evolves faster than $\ell = 0$.

\section{Experiment}

\begin{figure}[h]  
\centering
\includegraphics[width=0.9\linewidth]{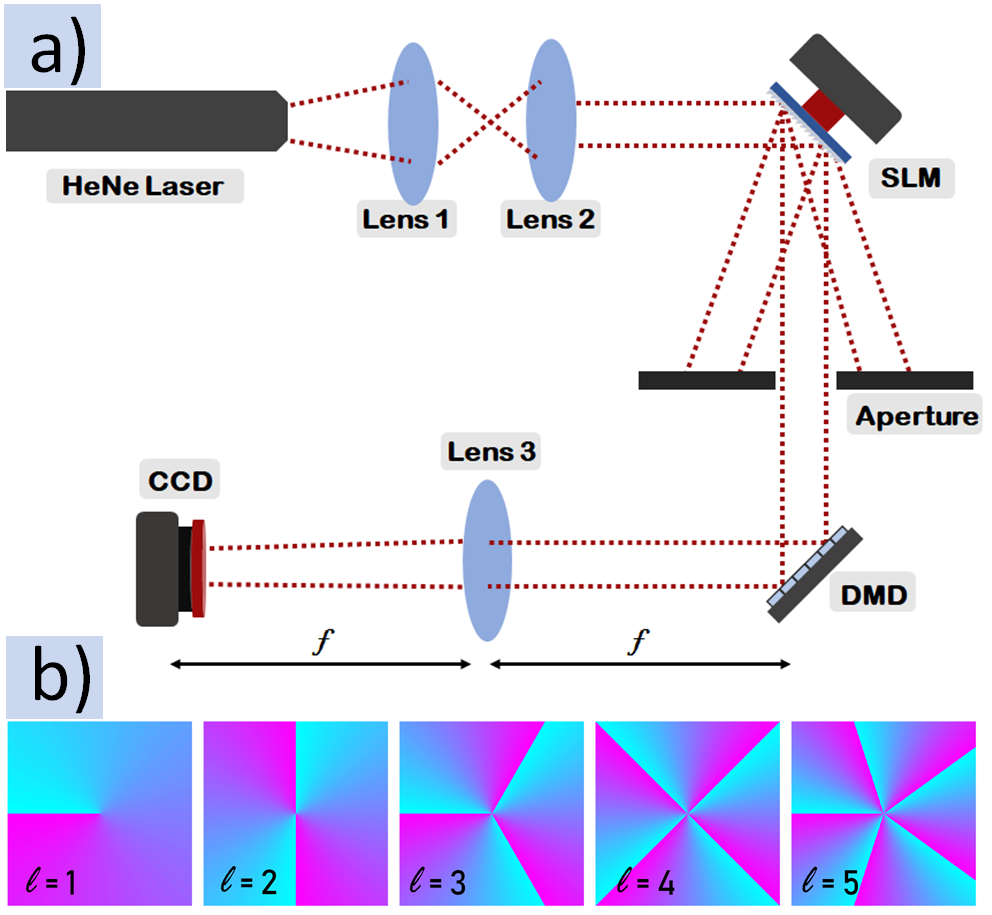}
\caption{Experimental setup (a). The LG beams were produced by uploading their phase in the SLM shown in (b). We were able to vary the position of the single slit by using a digital micromirror device (DMD) as a dynamic slit. The diffraction was imaged in the far-field using a CCD camera.}
\label{fig:setup}
\end{figure}

We generated the LG beam using the setup shown in figure \ref{fig:setup}a. A 632.8 nm HeNe laser was collimated by lenses L1 and L2. The beam was then converted into LG beams of various topological charges by uploading their corresponding phase profiles into the spatial light modulator. These phases are shown in figure \ref{fig:setup}b. A digital micromirror device (DMD) was then used as a dynamic slit. The DMD is composed of the an array of 608 by 608 mirrors of dimension 10.8 $\mu m$ and is controllable through a computer. The slit was moved from -1 mm (left) to 1 mm (right) of the beam center. The slit thickness was 54 $\mu m$. Finally, the diffraction was focused using lens L3 with focal length 25 cm and was captured using a CCD camera in the back focal plane of L3. The experimental results are discussed in the next section.

\section{Results and Discussions}
\begin{figure}[!h]
\centering
\includegraphics[width=0.9\linewidth]{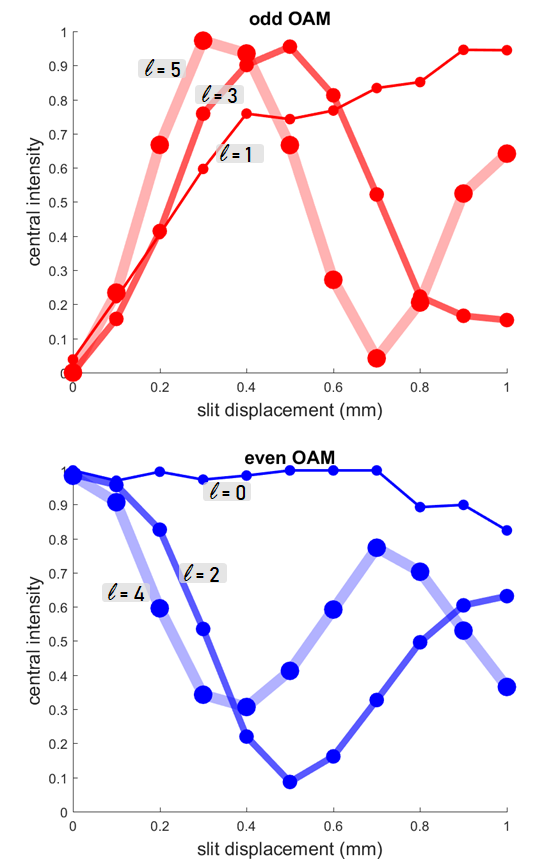}
\caption{Experimental result of central intensity vs position. Our results match our calculations and simulations that 1) as slit position is varied the central minima becomes a maxima for odd $\ell$, while the central maxima becomes a minima for even $\ell$ and that 2) higher $\ell$'s evolve faster than lower $\ell$'s.}
\label{fig:slitpositionexp}
\end{figure}

The intensity of the central peak as a function of the transverse displacement of the slit is shown in figure \ref{fig:slitpositionexp} and figure for odd and even $\ell$. The measured intensity for each position is the average of when the slit is a displaced to the left and the same distance to the right. For odd $\ell$, the intensity is zero when the slit is centered and increases to 1 as the slit is moved; while for even $\ell$, the opposite is true and the intensity is maximum when the slit is at the center and goes to zero as the slit is moved. These are all consistent with our calculations and our simulation in the previous section. One difference the we notice is that the experimental intensities do not reach the minima at zero. This is because we set the camera setting to high gain to capture our images specially for high $\ell$. We had to set the camera to increasing gain for higher $
\ell$ since the vortex is larger and the diffraction patterns becomes overall dimmer. However, although the minimum intensities do not completely reach zero, our eyes can still recognize them as intensity minima because the fringes are still visible.

In ref. \cite{ghai2009single}, the single slit diffraction of OAM beams was analyzed by taking the phase difference between the edge of the short axis of the slit. That is, the slit has some finite thickness. The analysis in ref. \cite{ghai2009single} essentially treats the single slit as a double slit. The result was that the fringes were not straight but are bent. In fact the same technique was used in \cite{sztul2006double} and in ref. \cite{emile2014young} for double slits and found the same bending of the fringes. Here, bending is not noticable since we used a very thin slit. For a very thin and long slit, the effect that there is a central minima and maxima for odd and even $\ell$ is more evident than the bending of the fringes.

In another paper, Ferreira et al \cite{ferreira2011fraunhofer} analyzed the diffraction of the LG beam through a single slit in terms of the phase difference between the end of the long axis of the slit. They explained the resulting pattern by approximating the LG beam by two separate Gaussian beams. Here we explain it by the method of stationary phase and considered only the phases at the endpoints of the slit. Recently, the method of stationary phase has been used to study sharp-edge diffraction of slits of various shapes\cite{narag2018diffraction,borghi2016catastrophe} . In this work, we also made use of a DMD as a dynamic slit in contrast with static slits in refs. \cite{ferreira2011fraunhofer,sztul2006double,emile2014young,ghai2009single}. We were also able to derive the relationship between the phase difference and the slit position in equation 5. In ref. \cite{ferreira2011fraunhofer} they also displaced the slit by some distance. However, they were only able to displace the slit to one certain position since they used a static slit. 

We note that the OAM can still be probed using stationary slits. However, it becomes harder and harder to resolve the fringes just by looking at the diffraction pattern, as in figure \ref{fig:DiffractionSimulation}. The number of horizontal fringes = $\ell + 1$ only works for small $\ell$, $\ell < 3$ in our case. In ref. \cite{ambuj2011far}, the number of fringes = $\ell + 1$ trend was also observed but only for $\ell > 3$. The difference between our method and that of ref. \cite{ambuj2011far}, is that they displaced the slit along the beam axis to observe the effect of the radial variation of the phase of the beam. Here we have displace the slit transversely and we have shown that displacing the slit allows us to access higher $\ell$. Our method, however, has a limitation. We cannot distinguish between the handedness of the OAM. We can only determine the absolute value of the OAM but not whether it is positive or negative. We also only investigate the diffraction for beams with integer $\ell$s. A future study in this field can look at beams with non-integer $\ell$.
\section{Conclusions}
We demonstrated that the OAM of an LG beam can be determined by looking at its diffraction through a single slit. On the basis of the method of stationary phase, we derived the difference between the endpoints of the slit we showed that for odd $\ell$ there is a central minima and for an even $\ell$ there is a central maxima. Furthermore, we also investigated the evolution of this phase difference as the slit is moved transversely. We used a digital micromirror device as a dynamic slit whose position can be controlled through a computer. We showed that although OAM of the slit can be determined using stationary slits, but with a dynamic slit we can access higher OAM values by looking at how the central horizontal fringes evolves as we change the slit position. We found that the evolution from intensity minima to maxima or maxima to minima is faster for larger $\ell$.

\begin{acknowledgments}
The authors acknowledge the Office of the Chancellor of the University of the Philippines Diliman through the Office of the Vice Chancellor for Research and Development Outright Research Grant for funding support. N. Hermosa is a recipient of the Balik PhD Program of the Office of the Vice President for Academic Affairs of the University of the Philippines. J. Narag is a Science Research Specialist of the PCIEERD Project No. 04002 under the Department of Science and Technology of the Republic of the Philippines.

\end{acknowledgments}

\bibliography{bibfile}

\end{document}